# A Transponder Aggregator with Efficient Use of Filtering Function for Transponder Noise Suppression

Kenya Suzuki, *Member, IEEE*, Osamu Moriwaki, *Member, IEEE*, Koichi Hadama, Keita Yamaguchi, Hiroki Taniguchi, Yoshiaki Kisaka, Daisuke Ogawa, Makoto Takeshita, Stefano Camatel, Yiran Ma, *Senior Member, IEEE*, Mitsunori Fukutoku

*Abstract*— Colorless, directionless, and contentionless reconfigurable optical add/drop multiplexing (CDC-ROADM) provides highly flexible physical layer network configuration. Such CDC-ROADM must operate in multiple wavelength bands which are being increasingly implemented in optical transmission systems. The operation in C+L bands requires switch devices used in CDC-ROADM to also be capable of multiband operation. Recent studies on wavelength division multiplexing (WDM) systems have pointed out the impact of amplified spontaneous emission (ASE) noise generated by signals of different wavelengths, which causes OSNR degradation. Therefore, it is desirable to filter out the ASE noise from different transponders when multiplexing multiple wavelengths at the transmitter side, especially in a system with non-wavelength selective combiners such as directional couplers and multicast switches. The use of transponder aggregators with filtering functions, such as the M × N wavelength selective switch (WSS), is preferable for this filtering. However, the downside of these devices is that it is difficult to provide economical multiband support. Therefore, we propose an economical transponder aggregator configuration by allowing a certain amount of ASE superposition and reducing the number of filtering functions. In this paper, we fabricated a prototype of the proposed transponder aggregator by combining silica-based planar lightwave circuit technology and C+L band WSS, both commercially available, and verified its feasibility through transmission experiments. The novel transponder aggregator is a practical solution for a multiband CDC-ROADM system with improved OSNR performance.

## I. INTRODUCTION

COLORLESS, directionless, and contentionless (CDC) reconfigurable optical add/drop multiplexing (ROADM) is widely used as a physical layer network architecture that enables the most efficient use of optical fiber, transmitters, and receivers in inter datacenter communications as well as in core and metro networks [1-10]. CDC-ROADM is used to allocate communication resources in response to changes in hot and cold spots in 5G communications and control load balancing in data center campuses. The following factors are important for meeting the expectations of such applications: (i) providing high transmission capacity over a wide geographic area, (ii) increasing the total transmission capacity handled by a ROADM network, (iii) providing a sufficient number of optical paths to ensure the high degree of freedom to change the network configuration, and (iv) economics. For (i), it is effective to increase the baud rate of the signal. That is, when comparing at the same bit rate, the signal format can be made less multi-leveled by increasing the baud rate of the signal. As a result, the noise tolerance of the signal can be increased, which leads to a greater transmission distance [11]. However, the baud rate expansion of the signal does not directly contribute to (ii), which is to increase the capacity of the signal that the network can handle. One possible solution is to utilize space division multiplexing (SDM) using multicore or multi-mode fibers [12-14]. However, at this point, networks utilizing multicore fibers have not been put into practical use. The variety of optical devices are not available, including transceivers suitable for SDM, and it will take time before these optical devices, including SDM fiber, can be developed economically. Other countermeasures to increase capacity include expanding the wavelength band used for optical communications, i.e., the use of multiband, which is considered an effective measure until the practical application of SDM fiber is feasible. To date, the use of multiple bands (C+L bands) from a single band (either C or L bands) has been demonstrated [15], and the use of S band has also been discussed [16-18]. Multiband usage is also important for (iii); it is effective to increase the number of signals provided by CDC-ROADM. The number of wavelength signals in wavelength division multiplexing (WDM) can be more than doubled by expanding the wavelength bands handled by the system. For example, when signals with a baud rate of 130 Gbaud is transmitted in a

K. Suzuki, O. Moriwaki, K. Hadama, and K. Yamaguchi are with NTT Device Innovation Center, NTT Corporation, Atsugi, Kanagawa 243-0198, Japan (e-mail: kenya.suzuki.mt@hco.ntt.co.jp).
Y. Kisaka and H. Taniguchi are with NTT Network Innovation Laboratories, NTT Corporation, Yokosuka, Kanagawa 239-0847, Japan.
D. Ogawa, M. Takeshita, and M. Fukutoku are with NTT Electronics Corporation, Yokohama, Kanagawa 221-0031, Japan.
S. Camatel and Y. Ma are with Finisar Corporation. Rosebery NSW 2018, Australia.

WDM system with a frequency spacing of 150 GHz, only about 30 signals can be allocated in the C-band (4800 GHz), whereas 60 signals can be arranged if the L-band is also utilized [15]. Flexible grids can also be used to increase the degree of freedom in networks [19] where signals with variety of baud rate are used. This approach may appear to be the opposite of the measure for the factor (i). However, increasing the number of signal channels while maintaining transmission capacity by increasing spectral efficiency by using a higher multilevel format and lowering the baud rate can also increase the degree of freedom of the network. The challenge of using the higher multilevel format is reducing the OSNR tolerance. In particular, when a large number of wavelength signals are transmitted through a WDM transmission line, the optical noise output by transponders of different wavelengths degrades the signal. The effect of this phenomenon is more pronounced as the number of signals increases. To address this issue, an optical filter (bandpass filter) should be inserted just after the transmitter to reduce amplified spontaneous emission (ASE) noise. The optical filter can also be installed in the transponder aggregator, which is a key device in the CDC-ROADM. The transponder aggregators are classified into two configuration. One is multicast switch [20, 21] and the other M × N wavelength selective switch (WSS) [22, 23]. For multicast switch, optical filters have been placed between the transmitter and a multicast switch [20]. The M × N wavelength selective switch (WSS) has integrated the filter function in itself. The requirements for (i) to (iii) change on a case-by-case basis depending on the size of the network and how it is used, but the requirements will need to be fulfilled economically (iv).

As a solution to the above set of problems, we have developed the concept of a transponder aggregator for a multiband network, which is required in the high baud rate signal era. The concept includes filtering function to reduce the impact of ASE noise from other transponders. In Section 2, we propose a configuration of a multiband transponder aggregator with an efficient filter function as a means of addressing the above issues, and explain its principles, advantages, and disadvantages. The proposed configuration is compared with conventional multicast switches and the M × N WSS. Section 3 presents the static characteristics of the proposed transponder aggregator constructed using a silica-based planar lightwave circuit (PLC) and multiband WSSs and the results of transmission experiments for 400 Gbps (DP-16QAM, 66 Gbaud) signals. Section 4 concludes the paper.

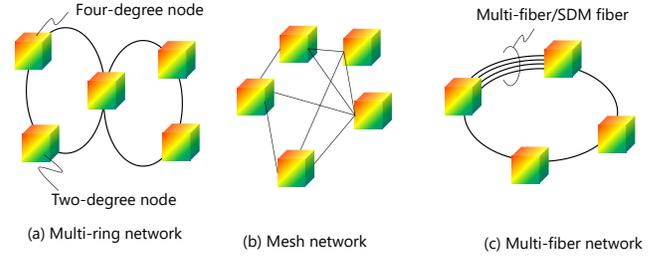

Fig. 1. Network configuration where CDC-ROADM nodes are allocated.

## II. TRANSPONDER AGGREGATOR WITH EFFICIENT USE OF FILTERING FUNCTION

### A. ROADM configuration and existing transponder aggregation technology

Figure 1 shows the multi-degree nodes to which CDC-ROADM is applied. These are located in the configurations of the networks in (a) the nodes that form ring-to-ring connections in the multi-ring network, (b), the multi-degree nodes in the mesh network, or (c) the multi-fiber nodes where multiple optical fibers are installed in the same direction. Transponders installed at such nodes must be utilized to be connected to any degree with a high degree of freedom. A coherent transponder has a tunable light source that can transmit and receive signals at any wavelength. In a node with such transponders, each transponder can transmit and detect optical signals at any wavelength and to/from any degree. This is called the colorless directionless (CD) function (Fig. 2 (a)). The equipment that connects these multiple transponders and optical cross-connects (OXCs) is called a transponder aggregator, and the transponder aggregator requires a function to connect a signal of an arbitrary wavelength to an arbitrary degree as described above. As shown in Fig. 2 (a), a simple M × N coupler can be used in CD-ROADM. In coherent optical communication, signals of different wavelengths are rejected by synchronous detection at the receiver to selectively receive the wavelengths. One of the drawbacks of CD-ROADMs is that the same wavelength cannot be used, even if the optical signals input to the transponder aggregator are directed to different degrees, because they are mixed in the M × N coupler. This constraint, called contention, limits the freedom of the optical network [2, 3].

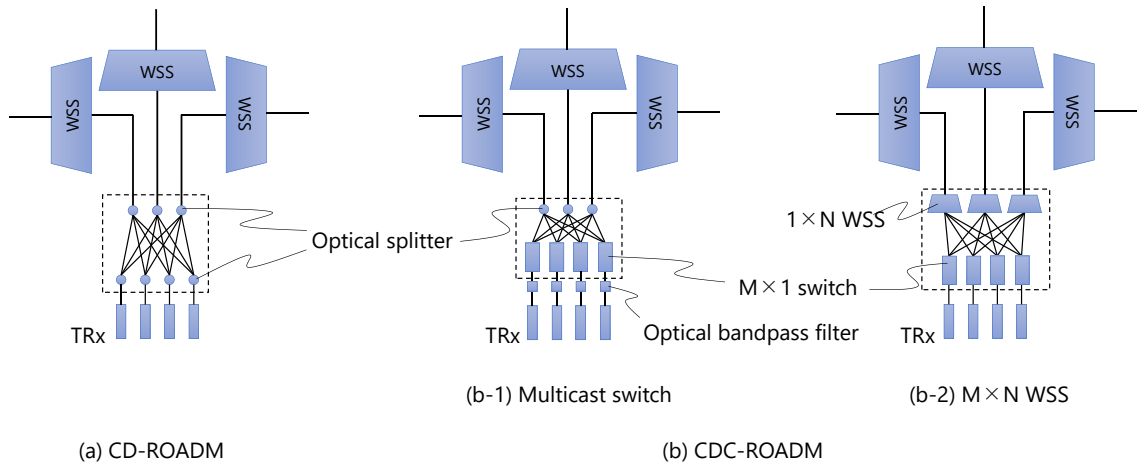

Fig. 2. Comparison of ROADM node configuration. (a) CD-ROADM node with M × N splitters, (b) CDC-ROADM with (b-1) multicast switch and (b-2) M × N WSS.

In contrast, the CDC-ROADM shown in Fig. 2 (b) implements a function that prevents contention in transponder aggregators. This figure shows typical CDC-ROADM node configurations. The contentionless transponder aggregators include an M × N multicast switch (Fig. 2 (b-1)) [20, 21] or an M × N WSS (Fig. 2 (b-2)) [22, 23]. The multicast switch has a braided configuration of N sets of 1 × M optical switches and M sets of N × 1 splitters. The optical signal from the transponder input to one of 1 × M switches is routed to the designated degree, multiplexed with other signals of different wavelengths by an N × 1 splitter connected to the degree, and routed to the OXC. Because only one optical signal can be transmitted to a transmission fiber connected to a certain degree for each wavelength, the same wavelength is never input to the splitter. Thus, signals from multiple transponders sent to a degree are all at different wavelengths. The multicast switch is made with a silica-based PLC or micro electro-mechanical system (MEMS). However, the N × 1 splitter is a passive element, which causes losses associated with multiplication and the degradation of OSNR due to ASE noise superimposed by other transponders of different wavelengths mixing into its own signal band.

The principle loss associated with the N × 1 splitter is

$$IL = -10\log\frac{1}{N}, \quad (1)$$

and the optical signal-to-noise ratio (OSNR) of the signal that input to the OXC is expressed by

$$OSNR = -10\log(10^{-\frac{OSNR_{in}}{10}} + (K-1) \times 10^{-\frac{OSNR_{out}}{10}}). \quad (2)$$

Here, $OSNR_{in}$ is the in-band OSNR of the signal itself at the wavelength, and $OSNR_{out}$ is the contribution of the noise generated by a signal other than the concerning signal. $K$ is the number of signals transmitted through the multicast switch to the degree. Thus, out-of-band noise of $K-1$ from transponders other than the signal band are added.

The M × N WSS was proposed to address the OSNR degradation occurring in multicast switches [22, 23]. In the configuration of the M × N WSS, M sets of N × 1 splitters in a multicast switch are logically replaced by M sets of N × 1 WSSs as shown in Fig. 2 (b-2). An advantage of this configuration is that, depending on the scale of the transponder aggregator, the insertion loss is lower than that of the MCS because the N × 1 WSS has no principle loss. In addition, the M × N WSS is also advantageous in terms of OSNR because the filtering function of the WSS can suppress the out-of-band noise. However, the M × N WSS is limited by its lack of scalability and multiband support, and additional filtering will result in a filtering penalty on the OSNR as well. Liquid crystal on silicon (LCOS) is currently the only commercially mature solution as a switching engine of a WSS, which implements the flexible grid function described in the introduction, but the figure of merit as a WSS is limited by the resolution and area of the LCOS panel. The figure of merit of the scalability of the M × N WSS can be expressed by the following formula [22]

$$M \times N \propto H \cdot \theta_{max} \quad (3)$$

where $H$ is the area or height of the LCOS panel and $\theta_{max}$ is the deflection angle achievable by the LCOS. The larger the deflection angle, the higher the required resolution of the panel. Because the M × N WSS requires M sets of 1 × N WSSs in each degree port, the number of transponders $N$ that can be connected inevitably decreases when trying to handle a large number of degrees $M$. To increase the figure of merit, LCOS with a large area and high definition must be used. However, the commonly available resolution of LCOS is about FHD (1920 × 1080), and the panel size should be small to keep costs down. As seen from eq. (3), a larger M × N WSS causes an increase in both the deflection angle and the panel area. This differs from display applications, which is the main target market of LCOS as its main applications are high-end projectors and head-mounted displays (HMDs). High-end projectors are expensive, and HMDs need to be lightweight and inexpensive, which are in contrast to the M × N WSS requirements.

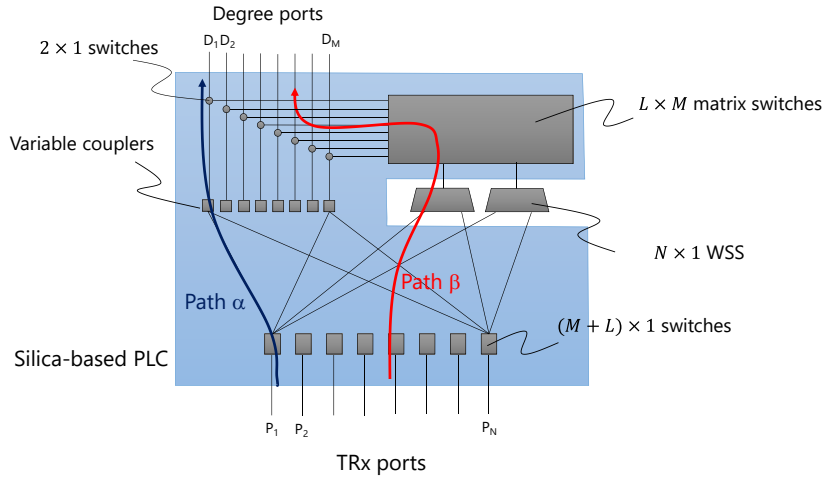

Fig. 3 Proposed configuration of novel transponder aggregator with efficient filtering function usage.

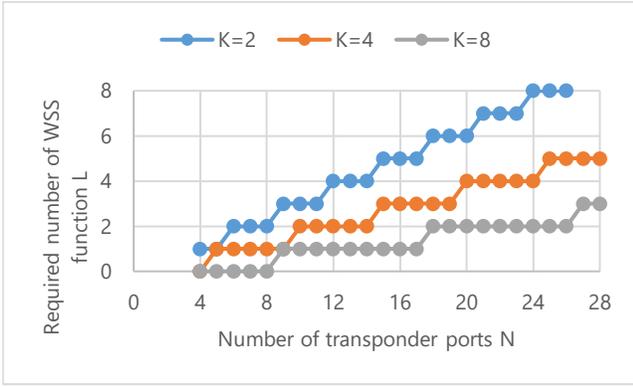

Fig. 4. Necessary number of WSS function versus number of transponders connected to transponder aggregator.

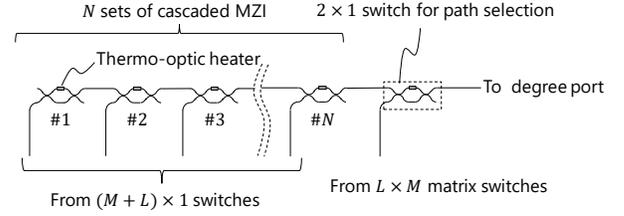

Fig. 5. Variable coupler configuration

*B. Device configuration and operation principle*

We propose a configuration for a transponder aggregator that effectively utilizes the filtering function with a minimal number of WSSs for out-of-band ASE suppression and reduces MCS loss at the same time. Figure 3 shows the proposed configuration for the M degree and N transponder. The blue area in the figure is the part that is configured as the integrated circuit of an optical waveguide such as a silica-based PLC, and conventional L units of 1 × N WSS are arranged outside. In the proposed configuration, the optical signal output from the transponder is input to any of N transponder ports and first travels to one of 1 × (M + L) optical switches. M ports of the 1 × (M + L) optical switch are connected to M sets of N × 1 variable splitters, and the remaining L ports are connected to expansion ports which are connected to one of N × 1 WSSs. The N × 1 variable splitter is connected to one input of the M sets of 2 × 1 switches. The common ports of the N × 1 WSSs are connected to one of the 2 × 1 switches via the L × M matrix switch integrated in the optical circuit.

The proposed transponder aggregator functions as follows. Inside the transponder aggregator, an input signal can be output to a degree port via two paths. The first path is indicated by α in Fig. 3, and the other is indicated by β. In path α, the optical signal is input through a 1 × (M + L) optical switch to an N × 1 variable splitter and routed through a 2 × 1 switch to one of the degree ports. No filtering is applied for this route. In the other path, β, the signal is routed by a 1 × (M + L) optical switch to one of the L units of N × 1 WSSs. After the out-of-band noise output from other transponders is filtered in the WSS, the desired output degree is selected by the L × M matrix switch and routed through the 2 × 1 switch connected to the designated degree.

What is noteworthy about this configuration is that the number of WSSs needed to be installed can be drastically reduced compared with the conventional configuration of M × N WSSs when the effect from out-of-band noise is allowed to a certain extent. In the proposed configuration, assuming that the number of signals to be aggregated via the first path α is $K$, or that the allowable number of out-of-band noise superimpositions is $K − 1$, the required number of N × 1 WSSs to be installed is represented by

$$L = \text{Int}\left(\frac{N}{K+1}\right). \qquad (4)$$

The equation is derived as follows. Because the proposed transponder aggregator allows out-of-band noise of $K − 1$ signal superimposition where $K$ signals are transmitted to a degree, the number of multiplexed signals that requires filtering function is $K + 1$ or more. These multiplexed signals to be filtered will be connected to the OXC via path β. If we consider the case where all $N$ signals from the transponders must pass through path β to be filtered, the number of WSSs required reaches the maximum when each WSS only needs to handle $K + 1$ signals. Therefore, the required number of WSSs can be derived by dividing the total number of transponders $N$ by $K + 1$, which leads to eq. (4). A relatively small numerical value is used to illustrate as an example. We assume a transponder aggregator to which eight transponders are connected ($N = 8$), and two signals are routed to degree port $D_1$ without filtering ($K = 2$). In this case, only the noise from a different wavelength is superimposed. The remaining six signals are output to a degree port other than degree port $D_1$, but because three signals are multiplexed when the WSS filtering function is required, it is sufficient to have two WSS functions (=Int (8/(2 + 1))) at most.

It should be noted that the number of WSSs required does not depend on the number of degrees but only on the number of transponders connected. In other words, the number of degrees can be expanded without implementing the WSS function for each degree as in the conventional M × N WSS, which greatly improves the efficiency of its usage. Figure 4 shows the relation between the number of transponders and the number of WSS functions required in the proposed configuration for $K = 2, 4,$ and 8. As the figure shows, even when 24 transponders are connected, there are eight WSSs when K = 2, which is then halved to four when K = 4, and quartered when K = 8.

The variable splitter in Fig. 3 consists of a series of Mach-Zehnder interferometers (MZIs) as shown in Fig. 5. An MZI is a common circuit element in silica-based PLC and is also used to construct multicast switches. The MZI consists of two directional couplers and two arm waveguides sandwiched between the couplers. By applying heat through a heater implemented on the top of one arm, the optical phase of signal through the arm is shifted through a thermo-optic effect so that the intensity of the output optical signal changes. By setting the MZI to an intermediate state rather than on/off, a variable coupler with an arbitrary splitting ratio is achieved. A serial connection of such MZIs results in a confluence of arbitrary number of signals. For example, when multiplexing the two signals from the transponder ports $P_1$ and $P_2$ in Fig. 3, MZI #1 and #2 can be set to 100:0 and 50:50, respectively. Meanwhile, when multiplexing the three signals from ports $P_1$, $P_2$, and $P_3$, they can be set to 100:0, 50:50, and 34:66, respectively. In conventional multicast switches, this variable coupler part is replaced by a fixed splitter of $1 \times N$, which causes principle loss due to splitting. For example, a multicast switch with eight transponders has a principle loss of 9 dB. However, by using the variable coupler in Fig. 5, the principle loss can be minimized with confluence. In the above case when $K = 2$, the principle loss is reduced to 3 dB. The variable coupler is suitable for fabrication by using waveguide technology. A MEMS is another candidate for multicast switches and variable optical attenuators. However, it is difficult to combine multiple signals while keeping intermediate states. The $1 \times (M + L)$ optical switches and the $L \times M$ matrix switch can be created using waveguide technology which involves many MZIs [24].

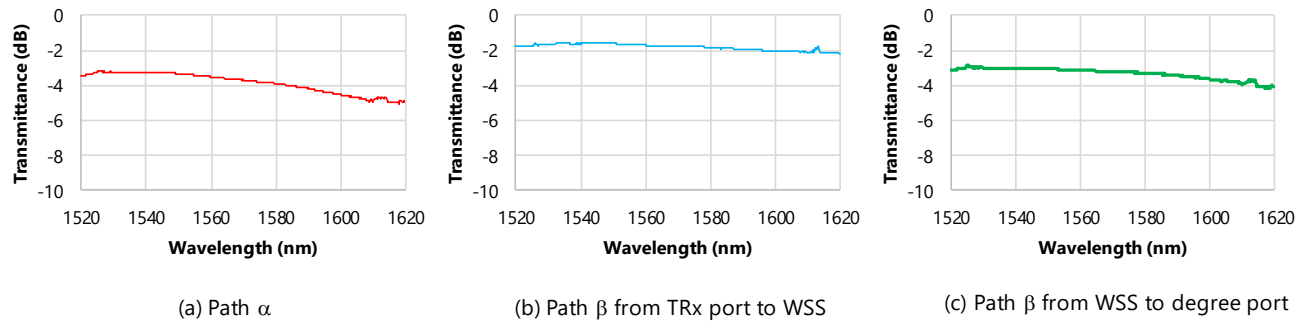

(a) Path α   (b) Path β from TRx port to WSS   (c) Path β from WSS to degree port

Fig. 6. Transmission spectra of fabricated transponder aggregator chip.

It should be noted that a drawback in the proposed configuration is that the degree of freedom degrades in certain reconfiguration scenarios. For example, when $N = 8, K = 2$ two transponders are connected to one degree port $D_1$ via route α. The remaining six transponders are connected to the other degree ports $D_2$ and $D_3$ via route β with three each, and one of the three signals connected to $D_2$ must be reconnected to $D_1$. The number of wavelengths connected to $D_1$ becomes three ($K = 3$), and a filtering function is required, so it should be connected via the WSS path β. Conversely, the WSS connected to $D_2$ should be used to connect to $D_1$ because the number of wavelengths is reduced from three to two. However, for this reconfiguration, signals that are already established must also be blocked to change the internal routing. For typical telecom use, this is not considered to be much of a problem because the amount of traffic in telecom is largely determined by degree. In addition, networks are rarely reconfigured, and CDC-ROADM is primarily used to remotely establish a new path with a newly installed transponder.

Another possibility is that a reroute may be needed due to a link down, such as a fiber break. For example, consider a case where four signals are connected to degree ports $D_1$ and $D_2$ with two each, and the remaining four signals are connected to degree port $D_3$. Suppose that the link at degree port $D_1$ is disconnected and the two signals connected to it must be rerouted through the link at $D_2$. Four signals will be transmitted to $D_2$, resulting in $K > 2$. In this situation, there must be at least one unused $1 \times N$ WSS because of eq. (4). The unused WSS can handle the two signals to be reconfigured through path β and have the $2 \times 1$ switch in Fig. 5 operate as a coupler rather than an on/off switch. In other words, signals originally connected to $D_2$ are routed through path α, and signals to be newly connected to $D_2$ are routed through path β, merged by the $2 \times 1$ switch acting as a variable coupler, and routed to $D_2$.

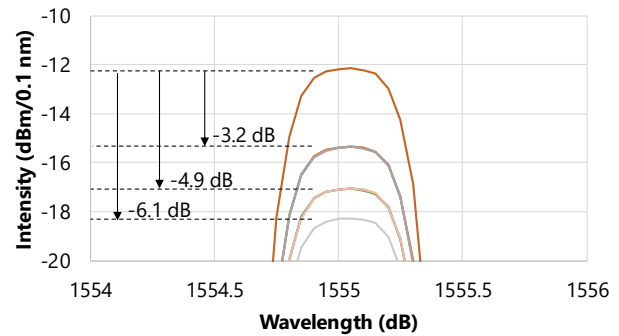

Fig. 7. Operation of variable coupler.

More simply, because a typical WSS used for OXC has 32 ports, multiple transponder aggregators may be installed in a node. In such cases, another option is to perform the reallocation by using other transponders connected to different transponder aggregators. Conversely, the blocking problem does not occur when a signal that is connected through an unfiltered path α is reallocated to a filtered path β.

III. EXPERIMENT

A. Static characteristics of proposed transponder aggregator

Using a silica-based PLC platform with a refractive index contrast of 2.5%, the transponder aggregator shown in Fig. 3 was fabricated with the number of degrees $M = 8$ and the number of connectable transponders $N = 8$. Figure 6 shows (a) a typical loss spectrum corresponding to path α, (b) a typical loss spectrum from a TRx port to the WSS function and (c) from the WSS function to the degree port in path β. For the measurements in Figure 6, the splitting ratio of the variable coupler was set to 100% for the measured paths. The results in Fig. 6 include fiber-to-PLC coupling losses and represent the excess loss of the fabricated transponder aggregator chip. The losses are broken down into fiber coupling losses of about 0.6 dB and losses due to errors in the coupling ratio of directional couplers in the MZIs from 50 %. The losses from the WSS to the degree port in the paths of Fig. 6 (a) and (c) were slightly larger because the optical signals experiences more MZI elements than the paths in Fig. 6 (b). Furthermore, the losses appear larger in the longer wavelengths due to design and manufacturing errors and can be easily minimized [24].

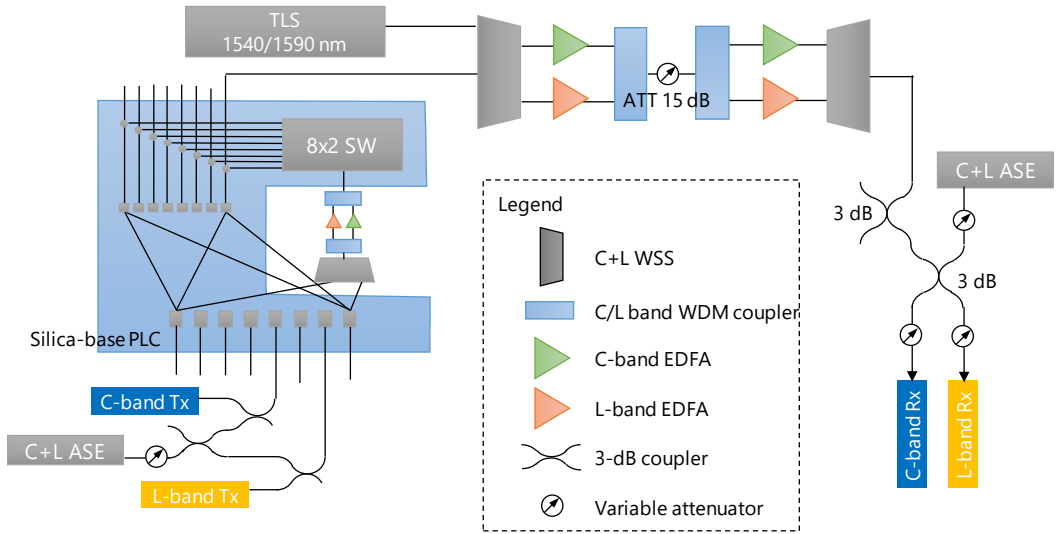

Fig. 8. Experimental setup for transmission experiment

Figure 7 shows the operation of the variable coupler. We input a 66 Gbaud DP-16QAM signal from a TRx port and measured the spectra at an output port. The plots show the cases when we set different sets of coupling ratios to the corresponding variable coupler. Namely, for single signal path connection, a 100 % coupling ratio was only set to MZI #1 in Fig. 5, and for two signal multiplexing, coupling ratios of 100 % and 50 % were set to MZIs #1 and #2, respectively. Similarly, for merging three signals, coupling ratios of 100 %, 50 %, and 33 % were set for MZIs #1, #2, and #3, and for merging four waves, coupling ratios of 100 %, 50 %, 33 %, and 25 % were set for MZIs #1, 2, 3, and 4. Thus, the coupling losses of the variable coupler are theoretically 3, 4.7, and 6 dB for two, three and four signal multiplexing, respectively. The obtained results in Fig. 7 are consistent with the theoretical losses, demonstrating the feasibility of the variable coupler.

*B. Transmission experiment*

We experimentally verified the applicability of the proposed transponder aggregator for transmission in ROADM nodes. In addition to reducing the number of WSSs, multiband operation is feasible with the proposed transponder aggregator. Figure 8 shows a schematic configuration of the experimental transmission system. We introduced C-band and L-band 400 Gbps signals (DP-16QAM, 66 Gbaud, PRBS 31 bit) into the proposed transponder aggregator and evaluated their bit error rates in the three scenarios described below. In the setup, the optical signal that was routed through the proposed transponder aggregator traveled through the C+L band WSS [25], which comprises the OXC at the add-node. Then the C-band and L-band erbium-doped fiber amplifiers (EDFAs) were used to compensate for the losses caused by the transponder aggregator and the OXC. After the C-band and L-band WSSs, a 15-dB loss was given by the attenuator, which simulates a transmission line. In addition, to operate both EDFAs, we introduced continuous-wave signals at wavelengths different from the signal wavelengths (1540 nm and 1590 nm) through other ports of the OXC so that the operating conditions for the minimum input power of the EDFAs were satisfied. After the propagation of the drop-side OXC, the signals experienced an emulated transponder aggregator with two 3-dB couplers connected as a loss medium. This configuration of the receiver side transponder aggregator is just for simplicity. As shown in the figure, a C+L band ASE was superimposed to simulate the noise given at the transmitter side and receiver side. The BER evaluation was carried out by setting the received signal power to −14 dBm.

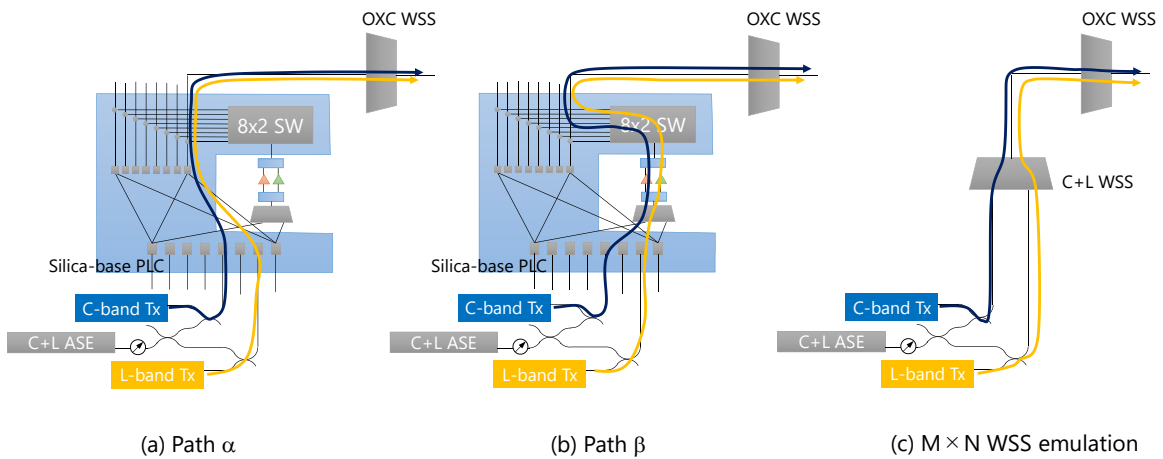

(a) Path α   (b) Path β   (c) M × N WSS emulation

Fig. 9. Scenarios in transmission experiment.

The following three scenarios were used to evaluate the bit error rate (BER) as shown in Fig. 9. In the first scenario, the signal travels through path α, for which noise from other transponders is superimposed. $K = 2$ and 4 were tested. The second scenario is a path of β. Here, only the transmitter OSNR (TOSNR) of the transmitted signal was considered because ASE noise from other transponders is suppressed by the WSS. In this scenario, as shown in the previous section, a loss of approximately 6 dB was expected in the fabricated transponder chip of the silica-based PLC and 5–7 dB in the WSS. Thus, a C+L band optical amplifier was installed at the common port of the WSS to compensate for these losses. Even if such an optical amplifier is installed, the number of amplifiers is at most $L$ sets, so the impact is limited. In addition, although we have multiple paths (combinations of TRx ports and degrees) for both paths α and β, only one combination was tested for simplicity. The third scenario is for a transponder aggregator simulating the M × N WSS, which was set up for comparison. In this scenario, the C+L WSS is considered a transponder aggregator. Again, only TOSNR was considered for the filtering function of the WSS. The third scenario is just for comparison because the C+L band M × N WSS has limited scalability. In these scenarios, BER was measured in accordance with the specification of OpenROADM [26] for reference. Namely, the TOSNR was set at 36 dB, and the out-of-band noise superimposed from other transponders was 43 dB. The TOSNR and out-of-band noise were emulated by adding equivalent ASE noise to the output signal from transponders. The transmission bandwidth of each WSS was also set to 87.5 GHz in accordance with the channel spacing specified in OpenROADM. Figure 10 shows the test signal spectra with a TOSNR of 36 dB. To check the multiband operation, we used the wavelengths of both edges and the center of the C- and L-bands, i.e., 1530.725, 1545.92, 1561.419, 1571.445, 1588.199, and 1605.314 nm.

Figure 11 shows the dependence of BER on the OSNR at the receiver when the signal is transmitted in the three scenarios. The six graphs illustrate the signal wavelengths 1530.725, 1545.92, 1561.419, 1571.445, 1588.199, and 1605.314 nm, respectively. Each graph plots four measurements. The filled plots represent the BER for path α, which corresponds to the first scenario; the circles are for $K = 2$ and the diamonds are for $K = 4$. The TOSNR was set following eq. (2), and the receiver OSNR (ROSNR) was calculated using a similar equation for the noise superimposed just before the receiver. The empty dots are the results corresponding to the second and third scenarios with the filtering function. The circle plot represents the second scenario with path β, and that of squares indicates the function corresponding to the M × N WSS for reference. As seen in the figure, C- and L-band multiband operation is feasible in the proposed configuration. The BER performance for the signal transmitted along path α overlapped

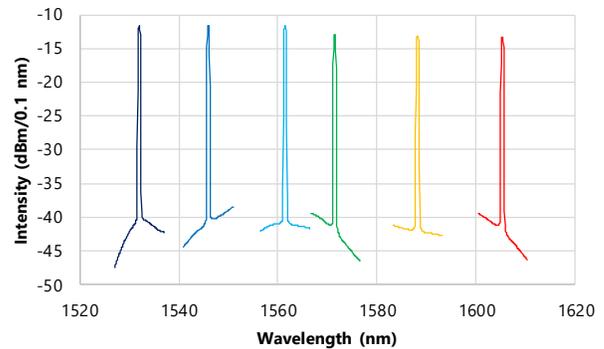

C-band: 1530.725 nm, 1545.92 nm, 1561.419 nm
L-band: 1571.445 nm, 1588.199 nm, 1605.314nm

Fig. 10. Test signal spectra

with that for the M × N WSS emulation of the third scenario. There is no observable degradation due to the use of the proposed transponder aggregator. The signal passing through path β appears slightly degraded compared with that of path α and the M × N WSS emulation. This is assumed to be caused by the EDFA installed inside the transponder aggregator. Because the effect of this EDFA occurs only once at the add-node, it is considered negligible when transmitting signals over long spans. For all wavelengths, the degradation of the ROSNR in path β from the other scenarios was less than 0.7 dB, giving a BER=$10^{-2}$. Overall, the BER of the C-band is lower than that of the L-band. This is due to the fact that the noise figures of the EDFA installed in the transmission path is large in the L-band than in the C-band. These results indicate that the proposed configuration does not cause degradation and that only the OSNR from the input to the output of the ROADM system needs to be considered.

As the results of this transmission experiment show, our proposed transponder aggregator configuration requires only a minimal integration of filter functions to reduce out-of-band ASE from other transponders.

IV. SUMMARY

In this paper, we proposed a configuration for transponder aggregators, which are a key element in CDC-ROADM. Out-of-band ASEs from transponders outputting different wavelength signals are known to degrade the OSNR at the transmitter side. The M × N WSS is expected to be effective in suppressing such degradation. However, the M × N WSS requires integrating as many WSS functions as the number of connected degrees, which is undesirable in terms of cost, because it requires a large LCOS panel for multiband operation. In our proposed configuration, the number of WSS functions for noise filtering is determined only by the number of transponders to be connected and not by the number of degrees. This results in much fewer functions than that of the conventional M × N WSS if the effect from out-of-band ASE is allowed to a certain extent. Our transponder aggregation device fabricated with a silica-based PLC operates in both C+L bands and demonstrates satisfactory static characteristics such as function of variable couplers. Transmission experiments using the proposed transponder aggregator combined with a C+L band WSS also showed that the performance of the proposed configuration is feasible. We believe that the proposed configuration is promising for the realization of multiband CDC-ROADM networks.

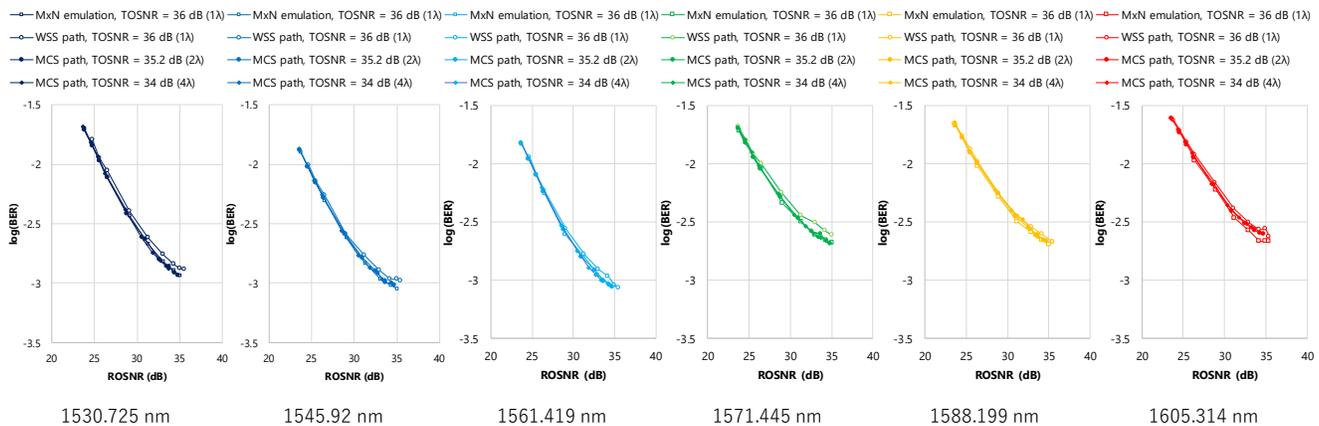

Fig. 11. Bit error rate versus receiver optical signal to noise ratio

**Kenya Suzuki** (M'00)
Kenya Suzuki received his B.E. and M.E. in electrical engineering and Dr. Eng. in electronics engineering from the University of Tokyo in 1995, 1997, and 2000, respectively. In 2000, he joined Nippon Telegraph and Telephone (NTT). From September 2004 to September 2005, he was a visiting scientist with the Research Laboratory of Electronics at the Massachusetts Institute of Technology, Cambridge, MA, USA. From 2008 to 2010, he was with NTT Electronics Corporation, where he was involved in the development and commercialization of silica-based waveguide devices. In 2014, he was a guest chair professor at the Tokyo Institute of Technology. He served as secretary of the Optoelectronics Research Technical Committee of the Institute of Electronics, Information and Communication Engineers (IEICE) of Japan from 2014 to 2016. He was a member of the subcommittee of Passive Optical Devices for Switching and Filtering of the Optical Fiber Communication Conference (OFC) from 2019 to 2021 and served as the chair of the subcommittee in 2022. He is currently a distinguished researcher and project leader at NTT Device Technology Laboratories. He is also the director of the New Energy and Industrial Technology Development Organization (NEDO) project on the photonic and electronic hybrid switch for the future datacenter networking of the Photonics Electronics Technology Research Association (PETRA). His research interests include optical functional devices, optical signal processing and automation of device design, and fabrication and evaluation of optical devices. He was the recipient of the Young Engineer Award and Electronics Society Activity Testimonial from IEICE in 2003 and 2016. He is a member of the Institute of Electrical and Electronics Engineers (IEEE), Optica, IEICE, and the Physical Society of Japan (JPS).

**Osamu Moriwaki**
Osamu Moriwaki received a B.E. and M.E. in electrical engineering from the University of Tokyo in 1998, and 2000, respectively. In 2000, he joined Nippon Telegraph and Telephone (NTT). He is an expert committee member of the technical committee on Photonic Network of the Institute of Electronics, Information and Communication Engineers (IEICE) of Japan. He is also the researcher of the New Energy and Industrial Technology Development Organization (NEDO) project on the photonic and electronic hybrid switch for future datacenter networking of Photonics Electronics Technology Research Association (PETRA). His research interests include optical switching devices and wavelength division multiplexed network system. He is a member of the Institute of Electrical and Electronics Engineers (IEEE), and IEICE.

**Koichi Hadama**
Koichi Hadama received a B.E. and M.S. in applied physics and advanced materials science from the University of Tokyo in 1999 and 2001. He joined the NTT Telecommunication Energy Laboratories, Atsugi, Kanagawa, Japan, in 2001. Since then, he has been engaged in the worked on research and development of free-space optical modules for WDM network systems. He is currently with the NTT Device Innovation Center, where he is engaged in development of optical and electrical modules for metro and access networks. He is a member of the Institute of Electronics, Information and Communication Engineers (IEICE).

**Keita Yamaguchi**
Keita Yamaguchi received B.S. and M.S. degrees in physics from Tsukuba University, and a Dr. Eng. degree in Computational Science and Engineering from Nagoya University in 2009, 2011 and 2019, respectively. In 2011, he joined Nippon Telegraph and Telephone (NTT) Corporation. He is currently with NTT Device Innovation Center, Atsugi, Japan. His research interests include silica-based waveguides. Dr. Yamaguchi is a member of the Institute of Electronics, Information and Communication Engineers (IEICE).

**Hiroki Taniguchi**
Hiroki Taniguchi was born in Kanagawa, Japan. He received B.E. and M.E. degrees in electrical engineering from the Tokyo Institute of Technology, Tokyo, Japan, in 2014 and 2016, respectively. He is currently a Research Engineer with Network Innovation Laboratories, Nippon Telegraph and Telephone (NTT) Corporation, Yokosuka, Japan. His research interest includes photonic transport network systems. He is a member of the Institute of Electronics, Information and Communication Engineers (IEICE). He was the recipient of the 2019 IEICE Communications Society Optical Communication Systems Young Researchers Award and the Young Researcher's Award from IEICE in 2020. He is a member of the Institute of Electronics, Information and Communication Engineers (IEICE).

**Yoshiaki Kisaka**
Yoshiaki Kisaka received B.S. and M.S. degrees in physics from Ritsumeikan University, Kyoto, Japan, in 1996 and 1998, respectively. In 1998, he joined NTT Optical Network Systems Laboratories, Yokosuka, Kanagawa, Japan, where he was engaged in research and development on high-speed optical communication systems including 40-Gbit/s/ch WDM transmission systems and the mapping/multiplexing scheme in the Optical Transport Network (OTN). From 2007 to 2010, he was with NTT Electronics Technology Corporation, where he was engaged in the development of the 40/100-Gbit/s OTN framer large-scale integration LSI. His current research interests are high-speed and high-capacity optical transport networks using digital coherent technology. Since 2010, he has contributed into the research and development of the digital coherent signal processor (DSP) for 100 Gbit/s and beyond. He is a member of the Institute of Electronics, Information and Communication Engineers (IEICE).

**Daisuke Ogawa**
Daisuke Ogawa is an Engineering Assistant Manager at NTT Electronics Corporation. He joined NTT Electronics in 1997. For the first two years, he was involved in the development of a silicon process. In 1999, he started working on silica-based planar lightwave circuit (PLC) devices and was engaged in the development and commercialization of temperature-insensitive AWGs, optical switches, ROADMs, and MCSs. In 2011, he joined NEL America, a subsidiary of NTT Electronics, where he spent a year as a sales representative for optical components. He is currently a group leader in the PLC department.

**Makoto Takeshita**
Makoto Takeshita received B.E. and M.E. degrees in photonics engineering from the University of Electro-Communications, Tokyo, Japan, in 2016 and 2018, respectively. His research was on an ultrafast semiconductor mode-locked laser for gas sensing and optical communications. In 2018, he joined NTT Electronics Corporation, where he has been engaged in the development of silica-based planar circuit (PLC) devices such as arrayed-waveguide gratings (AWGs), optical switches including multicast switches (MCSs) as a lightwave circuit designer and team lead engineer.

**Stefano Camatel**
Stefano Camatel received a Laurea degree in Electronic Engineering (summa cum laude) in 2001 and Ph.D. in Electronic and Communication Engineering in 2005, both from Politecnico di Torino. From June 2003 to June 2004, he was a visiting researcher at the University of California, Santa Barbara (UCSB). In 2005, Dr. Camatel was a postdoc at Politecnico di Torino working on new modulation formats for optical networks. From 2006 to 2008, he was a researcher at the Istituto Superiore Mario Boella working on free space optical communications, coherent detection, and plastic optical fiber communication systems. From 2008 to 2012, he worked as an R&D engineer and project manager with Nokia Siemens Networks in Munich, Germany. Since 2012, he has been working in product management for wavelength management products at Finisar Australia.

**Yiran Ma**
Yiran Ma received his master's and Ph.D. degrees in Telecommunication Engineering from The University of Melbourne, Australia, in 2006 and 2010. From 2010 to 2017, he was a Research Engineer with China Telecom Corporation Beijing Research Institute. He joined Finisar Australia as a system architect in 2017. His research interests include optical wavelength management, optical coherent communication and 5G transport networks. Dr. Ma is senior member of IEEE.

**Mitsunori Fukutoku**
Mitsunori Fukutoku received a B.E. and M.S. in electrical engineering from Tokushima University in 1989 and 1991, respectively. He joined NTT Transmission System Laboratories, Yokosuka, Kanagawa, Japan, in 1991. Since then, he has been engaged in the research and development of DWDM systems and ROADM systems. From 2000 to 2003, he was involved in optical transport network planning at NTT Communications, building advanced DWDM networks and Raman-amplified DWDM networks. Currently, he is engaged in sales and marketing strategies for optical communication devices at NTT Electronics. He is a member of the Institute of Electronics, Information and Communication Engineers (IEICE).